\documentclass[aps,pra,twocolumn,reprint,noeprint,superscriptaddress,amsmath,amssymb]{revtex4-1}
\usepackage{graphicx}
\usepackage{bm}
\usepackage{color}
\usepackage{subfigure}
\usepackage{bbm}
\usepackage{amsmath}
\usepackage[colorlinks, linkcolor=blue, urlcolor=blue, citecolor=blue]{hyperref}
\usepackage[toc,page]{appendix}

\begin{document}
\title{Cyclic nonlinear interferometry with entangled non-Gaussian spin states}
\author{Qi Liu}
\email[]{qi.liu@lkb.ens.fr}
\affiliation{Laboratoire Kastler Brossel, Coll\`ege de France, CNRS, ENS-PSL University, Sorbonne Universit\'e, 11 Place Marcelin Berthelot, 75005 Paris, France}
\author{Tian-Wei Mao}
\affiliation{State Key Laboratory of Low Dimensional Quantum Physics, Department of Physics, Tsinghua University, Beijing 100084, China}
\author{Ming Xue}
\affiliation{College of Physics, Nanjing University of Aeronautics and Astronautics, Nanjing 211106, China}
\author{Ling-Na Wu}
\email[]{lingna.wu@hainanu.edu.cn}
\affiliation{Center for Theoretical Physics and School of Science, Hainan University, Haikou 570228, China}
\author{Li You}
\email[]{lyou@mail.tsinghua.edu.cn}
 \affiliation{State Key Laboratory of Low Dimensional Quantum Physics, Department of Physics, Tsinghua University, Beijing 100084, China}
 \affiliation{Frontier Science Center for Quantum Information, Beijing 100193, China}

\date{\today}


\begin{abstract}
We propose an efficient nonlinear readout scheme for entangled non-Gaussian spin states (ENGSs) based on the intrinsic quasi-cyclic dynamics of interacting spin-1/2 systems. We focus on two well-known spin models of twist-and-turn (TNT) and two-axis-counter-twisting (TACT), where ENGS can be generated by spin dynamics starting from unstable fixed points. In the TNT model, non-Gaussian probe state evolves directly back to the vicinity of initial state during the subsequent time-forward evolution for path recombining, accompanied by quantum magnification of encoded signal and refocusing of the associated quantum noise. Based on low-order moment measurement, we find the optimal metrological gain nearly saturates the quantum Cram{\' e}r-Rao bound (QCRB) and follows the Heisenberg scaling. For the TACT case, the QCRB can also be nearly approached when the state converges to either of the two unstable fixed points, respectively corresponding to the initial state or its orthogonal coherent state. The latter case goes beyond previous studies where tracing back to or crossing the initial states is mostly considered. The present protocol does not require time-reversal as in typical nonlinear interferometries, and it also avoids complicated measurement of nonlinear observables or full probability distributions. The operational approach we discuss presents a practical way for realizing high-precision and detection-noise-robust quantum metrology with ENGS.
\end{abstract}
\maketitle

\section{Introduction}
Mesoscopic spin ensembles play essential roles in quantum science and technology, where a useful quantum resource concerns many-body spin states with entanglement or non-classical correlations~\cite{Pezze2018}. One of the most studied type
is spin squeezed state~\cite{squeezedmilestone, Wineland:1992vb}, which has found potential applications in various fields, including detection of electromagnetic field~\cite{Ockeloen:2013aa, Bao:2020aa, A.:2021aa}, optical atomic clock~\cite{Pedrozo-Penafiel:2020aa} and characterization of nonlocal correlations~\cite{Schmied:2016aa} or multi-partite entanglement~\cite{Kunkel:2018aa, Fadel:2018aa}. {For quantum enhanced sensing, sensitivity beyond the standard quantum limit (SQL) has been achieved with spin squeezing~\cite{Gross2010, Riedel2010, Hosten2016, Braverman:2019aa}. However, it is widely recognized that the metrological potential, or quantum Fisher information (QFI) of spin squeezed state is shy of providing the theoretical optimum known as the Heisenberg limited precision~\cite{Pezze2018}.} {Long-term spin dynamics, on the other hand, can give rise to highly entangled non-Gaussian states (ENGSs) characterized by non-Gaussian spin distributions~\cite{Dubost:2012aa}, such as Greenberger–Horne–Zeilinger state~\cite{Chalopin:2018aa, Song:2019aa, Omran:2019aa}, over-squeezed state~\cite{Strobel:2014ux, Evrard:2019aa,Liu2021, Colombo:2022tz}, twin-Fock and Dicke state~\cite{Twin_Matter,Luo620, Zou6381, Lange:2018aa}. Such ENGS may promise higher QFI than the {Gaussian squeezed state}, therefore enabling higher metrological sensitivity~\cite{Macri:2016aa, Baamara:2021aa, Gessner:2019aa}. Efficient phase estimation typically requires the measurement of full probability distribution~\cite{Strobel:2014ux, Evrard:2019aa} or high-order squeezing parameters~\cite{Gessner:2019aa, Xu:2022aa}, both of which add to complications of experiments, and are also susceptible to detection noise.

In recent years, interaction-based readout has been proposed as a powerful approach to achieve Heisenberg limited metrological gain without stringent requirement on particle number detection resolution~\cite{Yurke:1986aa, Gabbrielli2015, Macri:2016aa, Davis, Frowis2016, Manceau:2017uq, Huang:2018ur, Huang:2018aa, TACT_magnification, IBR_spin, Szigeti, Schulte2020ramsey, TNT_magnification, Nolan}. In such an approach, time-reversed nonlinear dynamics is often employed to evolve back the very one for generating entangled probe state, leading to amplification of an encoded signal and achieving sensitivity beyond SQL. However, experimentally reversing many-body dynamics is challenging, thus most previous experiments in quantum optics~\cite{Hudelist2014}, Bose-Einstein condensates~(BECs)~\cite{Linnemann:2016aa, spin_nematic}, trapped ions~\cite{Burd, A.:2021aa} and cold thermal atomic systems~\cite{Hosten} are limited to time-reversed dynamics in the short-term Gaussian regime, with time-reversal effectively engineered with linear spin rotations or echoes. While demonstrating robustness to detection noise, applications of ENGS for reaching Heisenberg limit remain to be reported. Two recent experiments respectively in spin-1 BEC~\cite{Liu2021} and cold atoms in optical cavity~\cite{Colombo:2022tz} have reported progresses inching towards this direction. 
The former employs time-forward nonlinear readout based on quasi-cyclic spin-mixing dynamics to achieve sub-shot-noise phase sensitivity with ENGS, while the latter observes Heisenberg scaling of measurement precision from an exact sign-flip of Hamiltonian through controlling the detuning of cavity photons. 

In this work, we extend the original idea in Ref.~\cite{Liu2021} to more general cases of spin-1/2 systems when flipping the sign of many-body interaction is operationally impractical. We provide protocols to construct the cyclic nonlinear interferometer, and present theoretical studes of quasi-cyclic nonlinear dynamics in reading out the phase encoded in ENGS. We investigate the long-term spin dynamics perturbed by signal encoding after the entangled non-Gaussian probe state is generated. In the absence of the phase encoding perturbation, quasi-cyclic dynamics shall bring the ENGS to the immediate vicinity of a coherent spin state and refocuses the associated quantum noise, while a finite encoded signal leads to a remarkably different output state. The signal-to-noise ratio (SNR) is magnified at the end of nonlinear dynamics, thus giving rise to sensitivity beyond SQL. 
\begin{figure}[t]
	\centering
	\includegraphics[width=1\linewidth]{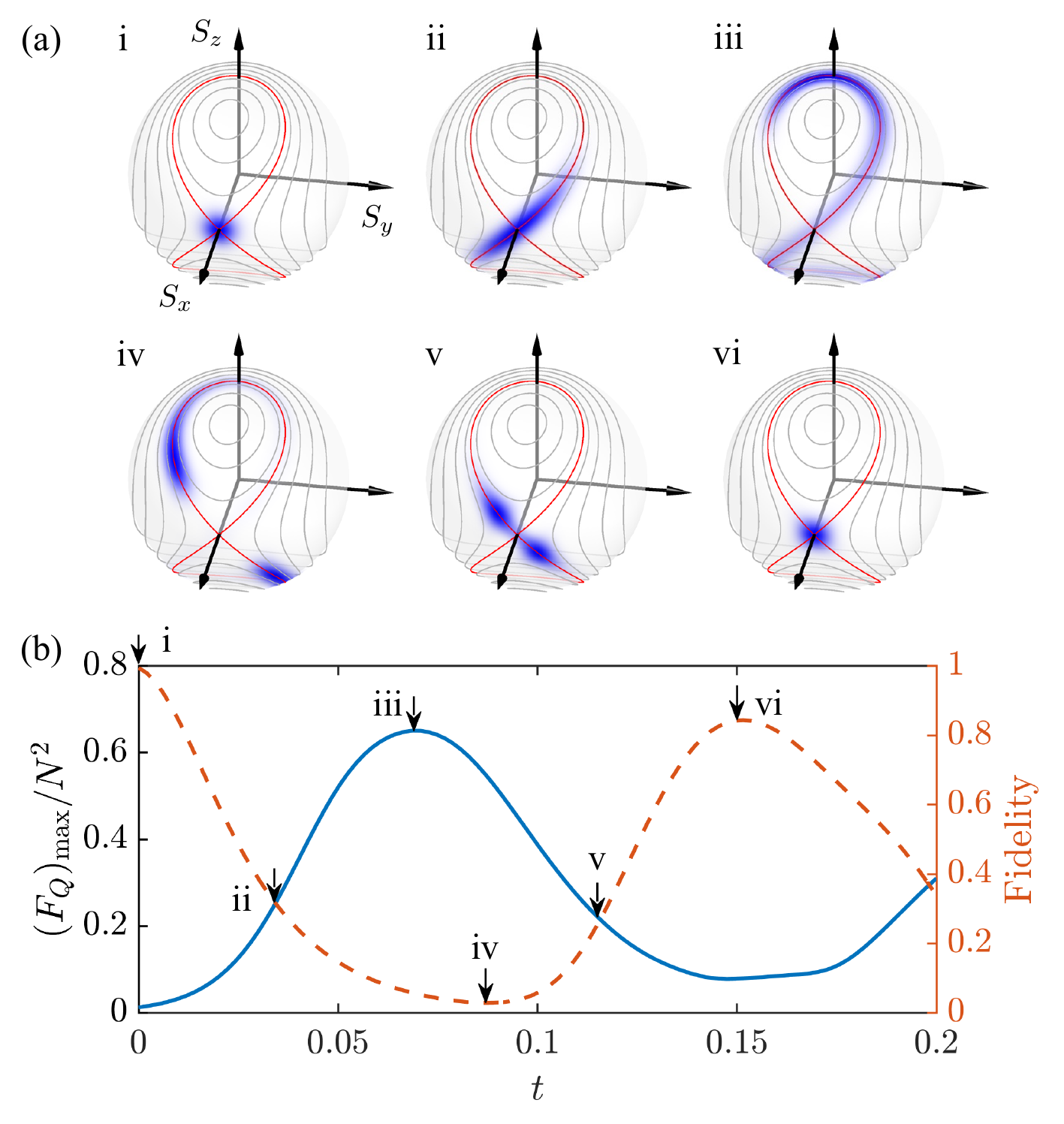}
	\caption{Quasi-cyclic dynamics in the TNT model. (a) Husimi representations of states during spin dynamics. Coherent state (i) prepared in the $\hat{S}_x$ eigenbasis $
\langle\hat{S}_x\rangle=N/2$ evolves into squeezed state (ii), ENGSs (iii-v), and back to the vicinity of initial point (vi) in succession. Solid curves denote the mean-field energy contours with the separatrix highlighted in red color. (b) Evolution of the normalized QFI maximized over phase encoding directions (blue line) and fidelity with respect to the initial state (red dashed line). The simulations are carried out with $N=100$, $\chi=1$, and $\Lambda=2$.}
 \label{fig1}
\end{figure}
This paper is organized as follows. In Sec.~\ref{TNT}, the twist-and-turn (TNT) model~\cite{Zibold:2010aa, Twist-and-turnexp, Bohnet:2016aa} is studied first. We describe the intrinsic quasi-cyclic dynamics in Sec.~\ref{TNT1}, and show the corresponding configuration of nonlinear cyclic interferometry in Sec.~\ref{TNT2}.
Section ~\ref{TNT3} introduces an analytical optimization method~\cite{Gessner:2019aa, Schulte2020ramsey} supporting for obtaining the {optimal rotation axis for signal encoding and measurement basis}. In Sec.~\ref{TNT4} and Sec.~\ref{TNT6}, we study the metrological performance of the nonlinear readout protocol, including the Heisenberg scaling of sensitivity with particle number and robustness to detection noise.
In Sec.~\ref{TACT}, we move on to the cyclic interferometer for the two-axis-countertwisting (TACT) model~\cite{squeezedmilestone, Kajtoch:2015aa, TACT_magnification}, where a new quantum magnification scheme is revealed without requiring explicitly tracing back to the initial state. A brief summary is  given in Sec.~\ref{conclusion}.


\section{Cyclic nonlinear interferometry in the TNT model}\label{TNT}
\subsection{Spin dynamics of the TNT Hamiltonian}\label{TNT1}
We first consider the well-known TNT Hamiltonian~\cite{Zibold:2010aa, Twist-and-turnexp, Bohnet:2016aa}
\begin{equation}
\hat{H}_{\rm{TNT}}=\chi\hat{S}_z^2+\Omega\hat{S}_x,
\end{equation}
which describes a spin-1/2 system with all-to-all coupling in the presence of linear transverse field. Here, $\hat{S}_i=\sum_{k=1}^N\hat{\sigma}_i^{(k)}/2~(i=x,y,z)$~($\hbar=1$ hereafter) denotes the collective spin operator with $\hat{\sigma}_i^{(k)}$ the Pauli matrix of the $k$-th particle, while $\chi$ and $\Omega$ stand for the strength of nonlinear interaction and transverse field, respectively. 

{Depending on the value of parameter $\Lambda=\chi N/\Omega$, the system exhibits two types of dynamics~\cite{Zibold:2010aa}}. In the Rabi oscillation regime where $\Lambda<1$, the linear coupling dominates the dynamics, thus spin evolution takes on regular periodic Rabi-like oscillations. While in the Josephson regime ($\Lambda>1$), there appears a dynamically unstable point at $\langle\hat{\mathbf{S}}\rangle=(\langle\hat{S}_x\rangle, \langle\hat{S}_y\rangle, \langle\hat{S}_z\rangle)=(N/2,0,0)$ [see Fig.~\ref{fig1}(a)i]. For the short-term evolution, coherent spin state prepared at this unstable point stretches along one branch of separatrix (marked by red mean-field energy contour) while shrinks along the other, i.e., evolves into a spin squeezed state [Fig.~\ref{fig1}(a)ii]. As the dynamics proceeds, the Husimi distribution of the state starts to wind around the surface of Bloch sphere [Fig.~\ref{fig1}(a)iii]. Such a state is known as over-squeezed state or non-Gaussian state~\cite{Strobel:2014ux}, due to its non-Gaussian distribution. Although spin squeezing is degraded for ENGS, the QFI maximized over signal encoding operator $\hat{S}_n$, given by $(F_Q)_{\rm max}=\max_{\mathbf{n}}4(\Delta\hat{S}_n)^2$~\cite{comment1}, actually becomes larger [see blue curve in Fig.~\ref{fig1}(b)], which therefore promises improved measurement precision lower-bounded by QCRB~\cite{Pezze2018} of $(\Delta\phi)_{\rm{QCRB}}=1/\sqrt{F_Q}$.

The signal encoded into ENGS cannot be efficiently extracted through the widely employed readout scheme of measuring low-order moments~\cite{Gessner:2019aa}, but precision approaching QCRB can be realized with time-reversed nonlinear readout, such that the final state detected recovers almost Gaussian statistics without beyond second-order correlations~\cite{Davis, Frowis2016}. For the TNT system discussed, this recovery can be automatically realized if time-forward dynamics is allowed to go on indefinitely time-forward. As shown in Fig.~\ref{fig1}(a)iv-vi, the Husimi distribution flows on the Bloch sphere and converges towards the initial state. At the instant of quasi-period (marked by vi), the fidelity of the final state with respect to the initial one touches a maximal value of $0.84$ [red dashed line in Fig.~\ref{fig1}(b)], while $F_Q$ decreases to a local minimum, demonstrating quasi-cyclic spin dynamics. 
We will show in the following how such quasi-cyclic behavior can be employed to magnify the signal encoded and realize measurement precision beyond SQL for the ENGS.

\subsection{Cyclic nonlinear interferometer}\label{TNT2}
The cyclic nonlinear interferometry demonstrated earlier in the spin-1 BEC system~\cite{Liu2021} enables quantum enhanced precision. Here, the operational configuration is adapted to spin-1/2 system as sketched in Fig.~\ref{fig2}(a). With a coherent spin state as the input, the first stage of nonlinear dynamics $U(t_1)$ serves as nonlinear splitting leading to ENGS. After sensing a spin rotation $R_n(\phi)=e^{-i\phi
\hat{S}_n}$ around $\hat{S}_n=\hat{\textbf{S}}\cdot\textbf{n}$, the system undergoes a second stage of dynamics $U(t_2)$ as in effective nonlinear recombining. Augmented by a final spin rotation $R_m$ and subsequent population measurement, spin observable $\hat{S}_m=\hat{\textbf{S}}\cdot\textbf{m}$ can be determined. Both unit vectors $\mathbf{n}$ and $\mathbf{m}$ can be optimized to provide the highest measurement precision of $\phi$, as we detail in the next section. 

Figure~\ref{fig2}(b) shows the Husimi representations of the input state, ENGS, and output state in succession. Among them, the ENGS generated by $U(t_1)$ is found to wrap around the Bloch sphere [middle panel in Fig.~\ref{fig2}(b)], accompanied by a wide non-Gaussian probability distribution on the $\hat{S}_z$ eigenbasis $\{s_z\}$ [Fig.~\ref{fig2}(c) blue data]. A small rotation of $\phi=1/\sqrt{N}$ around $\hat{S}_y$ axis slightly modifies the distribution [Fig.~\ref{fig2}(c) red data] but leaves the state nearly indistinguishable based on measuring only $\langle\hat{S}_z\rangle$ and $\Delta \hat{S}_z$~\cite{comment2}. To illustrate this, we show the corresponding Gaussian distributions with the center at $\langle\hat{S}_z\rangle$ and width of $\Delta\hat{S}_z$ [dashed lines]. One can see that the two Gaussian distributions (with and without rotation) are hardly distinguishable. However, the situation becomes strikingly different in the presence of the nonlinear recombining $U(t_2)$, which brings the state back to the vicinity of the initial one in the absence of encoding signal [Fig.~\ref{fig2}(d) blue data]. Remarkable difference emerges between the two distributions with and without rotation, making it possible to distinguish the two states by measuring only $\langle\hat{S}_z\rangle$ and $\Delta\hat{S}_z$. More importantly, the small rotation of probe state now evolves into a magnified shift of $\langle\hat{S}_z\rangle$ [Fig.~\ref{fig2}(d) red data], as is quantitatively shown in Fig.~\ref{fig2}(e). Together with the reduced quantum noise in the vicinity of $\phi=0$ [Fig.~\ref{fig2}(f)], the quasi-cyclic nonlinear readout gives an improved SNR compared to a direct linear readout of ENGS. 


\begin{figure}[t]
	\centering
	\includegraphics[width=1\linewidth]{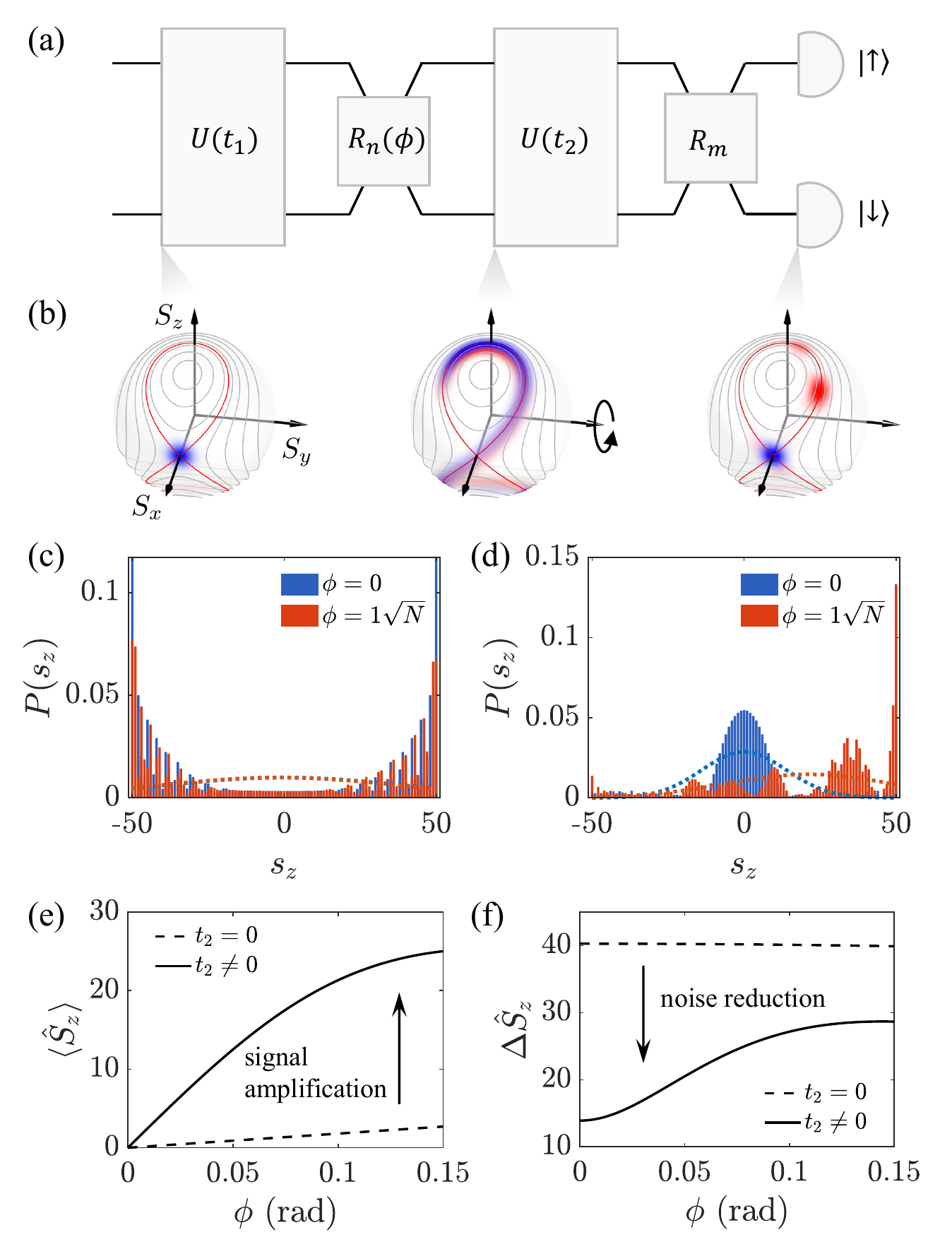}
	\caption{(a) The operational protocol for cyclic nonlinear interferometry in spin-1/2 systems,  consisting of nonlinear splitting $U(t_1)$, signal encoding rotation $R_n(\phi)$, nonlinear recombining $U(t_2)$, measurement rotation $R_m$, and population detection. (b) Husimi representations of the initial state, non-Gaussian probe state with (red shading) or without (blue shading) signal encoding and their corresponding output states. (c) Probability distributions of the probe states in $\hat{S}_z$ eigenbasis. Dashed lines denote the Gaussian fitting curves. (d) Probability distributions of the output states. Quasi-cyclic nonlinear readout refocuses the quantum noise of probe state while magnifies the shift of $\langle\hat{S}_z\rangle$ in the vicinity of $\phi=0$, thus making the encoded signal distinguishable with low-order moment measurement. (e) Dependence of signal ($\langle\hat{S}_z\rangle$) and (f) noise ($\Delta\hat{S}_z$) on the rotation angle $\phi$. Black dashed and solid lines refer to the cases without and with $U(t_2)$, respectively. We take $N=100$, $\chi=1$, $\Lambda=2$, $R_n(\phi)=e^{-i\hat{S}_y/\sqrt{N}}$, $R_m=1$, $t_1=0.07$, and $t_2=0.08$ in the simulations for illustration.}
 \label{fig2}
\end{figure}
\subsection{Optimizations of $\mathbf{n}$ and $\mathbf{m}$}\label{TNT3}
The attainable precision for cyclic nonlinear interferometry depends on the signal encoding operator $\hat{S}_n$ and observable $\hat{S}_m$, the former of which determines the QFI of a given probe state, thus sets a lower bound (or QCRB) to the ultimate measurement precision, while the latter one controls how closely we can approach this bound along a given measurement direction. 
 It is thus meaningful to optimize both $\mathbf{n}$ and $\mathbf{m}$ to obtain the highest metrological gain in theory. These two vectors can be determined analytically for given $U(t_1)$ and $U(t_2)$, as first introduced in~\cite{Gessner:2019aa} and then extended in~\cite{Schulte2020ramsey} for nonlinear readout. We now briefly introduce this method.

In Heisenberg picture, the observable $\hat{S}_m$ evolves into 
\begin{equation}
\hat{\tilde S}_m=U^{\dagger}(t_1)R_n^{\dagger}(\phi)U^{\dagger}(t_2)\hat{S}_mU(t_2)R_n(\phi)U(t_1)
\end{equation}
at the end of the nonlinear recombining operation. By defining $\hat{S}_m(U)=U^{\dagger}\hat{S}_mU$, we have 
\begin{equation}
\hat{\tilde S}_m=e^{i\phi\hat{S}_n(U(t_1))}\hat{S}_m(U(t_2)U(t_1))e^{-i\phi\hat{S}_n(U(t_1))},
\end{equation}
and therefore 
\begin{equation}
\partial\langle\hat{\tilde S}_m\rangle/\partial\phi=i\left\langle\left[\hat{S}_n(U(t_1)),\hat{\tilde S}_m\right]\right\rangle.
\end{equation}
It is convenient to re-express the equation in its matrix form \begin{equation}
\partial\langle\hat{\tilde S}_m\rangle/\partial\phi=\mathbf{n}^TM\mathbf{m},
\end{equation}
where $M$ is a $3\times3$ matrix with elements $M_{kl}=i\left\langle\left[\hat{S}_k(U(t_1)),\hat{\tilde S}_l\right]\right\rangle$ ($k,l=x,y,z$). In the same spirit, we determine the measurement noise 
\begin{equation}\label{variance}
(\Delta\hat{\tilde S}_m)^2=\langle\hat{\tilde S}_m^2\rangle-\langle\hat{\tilde S}_m\rangle^2=\mathbf{m}^TQ\mathbf{m},
\end{equation}
where $Q$ denotes the covariance matrix with elements $Q_{kl}=\langle\hat{\tilde S}_k\hat{\tilde S}_l+\hat{\tilde S}_l\hat{\tilde S}_k\rangle/2-\langle\hat{\tilde S}_k\rangle\langle\hat{\tilde S}_l\rangle$. Finally, the phase sensitivity can be obtained through the error propagation formula
\begin{equation}\label{psr}
(\Delta\phi)^{-2}=\dfrac{|\partial\langle\hat{\tilde S}_m\rangle/\partial\phi|^2}{(\Delta\hat{\tilde S}_m)^2}=\dfrac{(\mathbf{n}^TM\mathbf{m})^2}{\mathbf{m}^TQ\mathbf{m}}.
\end{equation} 

To find the maximal value of Eq.~\eqref{psr}, we define $\mathbf{v}=Q^{1/2}\mathbf{m}$ and $\mathbf{u}=Q^{-1/2}M^T\mathbf{n}$. According to Cauchy-Schwarz inequality, we obtain 
\begin{equation}\label{optimal_sensitivity}
(\Delta\phi)^{-2}=\dfrac{(\textbf{u}^T\mathbf{v})^2}{\mathbf{v}^T\mathbf{v}}\leq\mathbf{u}^T\mathbf{u}=\mathbf{n}^TMQ^{-1}M^T\mathbf{n},
\end{equation}
and the equality is achieved at $\mathbf{u}=\mathbf{v}$, or $\mathbf{m}=Q^{-1}M^T\mathbf{n}$. For the general case $\phi\neq0$, the momentum matrix $MQ^{-1}M^T$ depends on the selection of $\mathbf{n}$, which poses difficulties for maximizing the right-hand-side of the inequality (3). However, for the special case of $\phi=0$, the momentum matrix is solely decided by nonlinear dynamics $U(t_1)$ and $U(t_2)$, thus the lower bound of phase sensitivity $\Delta\phi$ is simply given by the maximal eigenvalue of $MQ^{-1}M^T$. The corresponding eigenstate gives the optimal signal encoding direction $\mathbf{n}_{\rm{opt}}$ and also the optimal measurement direction $\mathbf{m}_{\rm{opt}}=Q^{-1}M^T\mathbf{n}_{\rm{opt}}$. For simplicity, we restrict us to the latter case of $\phi=0$ in the following. This selection actually is sufficient to enable almost optimal measurement sensitivity as we will see.

\subsection{Time-dependence of metrological gain}\label{TNT4}
Given the finite life time of experimental systems, it is meaningful to balance the limited time resources among various stages of the interferometer (splitting, encoding and recombining) to optimize the sensitivity~\cite{Hayes:2018aa, Haine:2020aa}. In this part, we investigate the dependence of metrological gain on splitting time $t_1$ and recombining time $t_2$. The gain is defined as 
\begin{equation}
G=-20\log_{\rm{10}}\left[\Delta\phi/(\Delta\phi)_{\rm{SQL}}\right],
\end{equation}
in which $(\Delta\phi)_{\rm{SQL}}=1/\sqrt{N}$ denotes the SQL. For numerical convenience, we consider $N=100$, $\chi=1$, and $\Lambda=2$. For each $(t_1, t_2)$ combination, optimizations of $\mathbf{n}$ and $\mathbf{m}$ are adopted to maximize their metrological gain. 
\begin{figure}[t]
	\centering
	\includegraphics[width=1\linewidth]{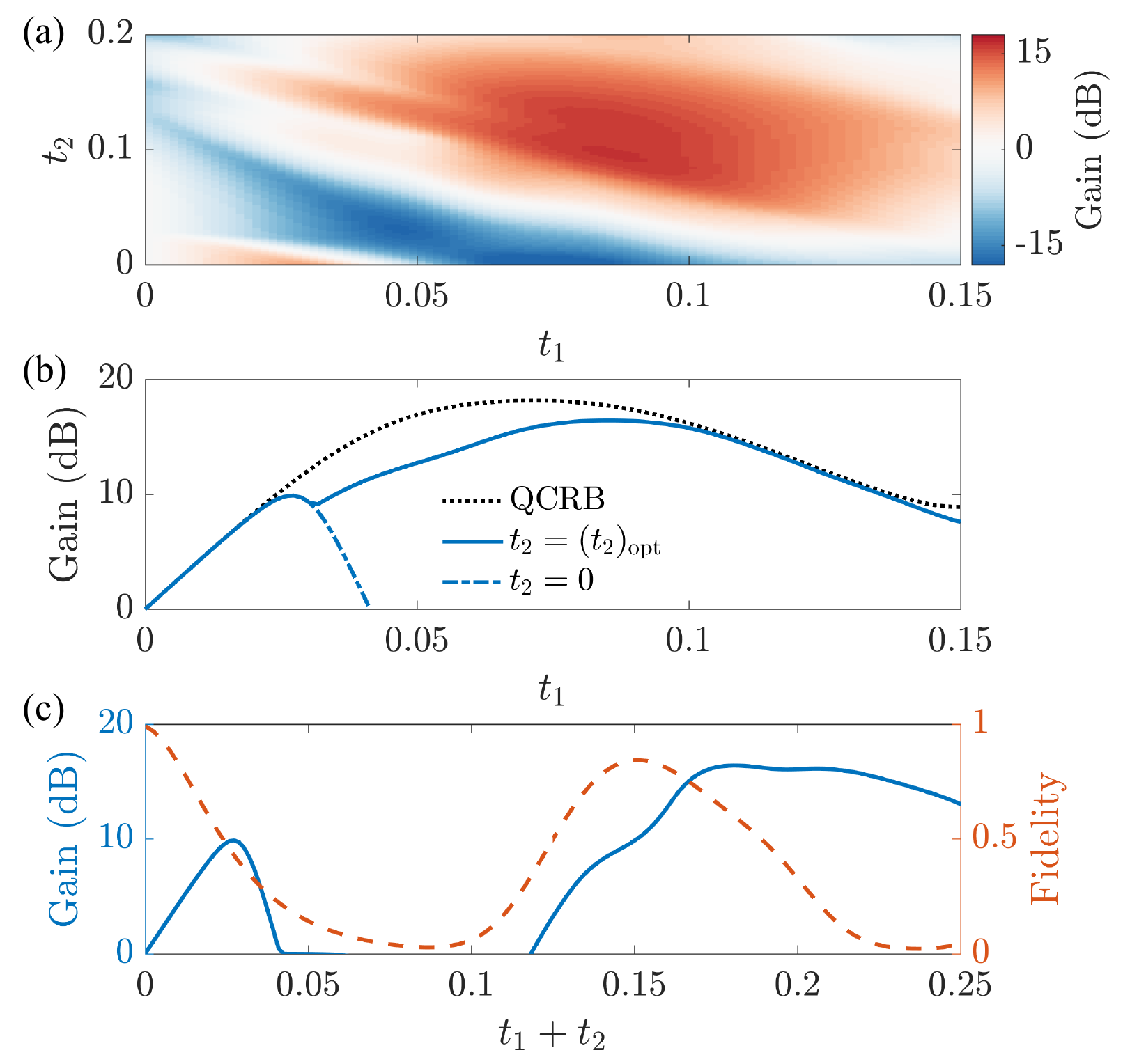}
	\caption{(a) Dependence of metrological gains on $t_1$ and $t_2$. (b) Gains optimized over $t_2$ for given $t_1$ (blue solid line). Blue dash-dotted line denotes the case of $t_2=0$, and black dotted line stands for the QCRB. (c) Optimal gain obtained at total time $t_1+t_2$ (blue line), and the corresponding fidelity with respect to the initial state (red dashed line). $N=100$, $\chi=1$, and $\Lambda=2$ are used in the simulations.}
 \label{fig3}
\end{figure}
As shown in Fig.~\ref{fig3}(a), we find sub-shot noise sensitivity ($G>0$, denoted by red color) can be achieved over a wide range of $t_1$ and $t_2$ values. {We can recognize two regions with positive gains in this 2D map. The first region dominates in the limit of $t_2\rightarrow0$, where positive gain can only be achieved within $t_1<0.05$.} In detail, we find the probe state in this regime corresponds to a Gaussian squeezed state, which can be directly read out without the help of nonlinear dynamics. However, with increasing $t_1$, the probe state becomes over-squeezed and the metrological gain starts to decrease, as is more clearly revealed by the blue dash-dotted line in Fig.~\ref{fig3}(b), which is calculated with $t_2=0$. The second region is characteristic of large total duration $t_1+t_2$. Higher gains are obtained thanks to the ENGS generated at larger $t_1$. We note at $t_1=0.085$, the metrological gain reaches the maximal value of $16.4~\rm{dB}$, which is only slightly below the largest QCRB of $18.1~\rm{dB}$ {expected at $t_1=0.07$}. The deviation from QCRB can be ascribed to the imperfect disentanglement of the quasi-cyclic dynamics, which leaves the final state non-Gaussian. We show in Appendix~\ref{AppA} that the measurement of higher-order collective spin operators~\cite{Gessner:2019aa} can lead to larger metrological gains and a broader parameter regime where QCRB is saturated.

To provide a quantitative understanding of the role played by quasi-cyclic dynamics, we plot the optimal gain achieved with total duration of $t_1+t_2$ in Fig.~\ref{fig3}(c). The gain curve (blue solid line) is clearly composed of a Gaussian ($t_1+t_2<0.05$) and a non-Gaussian region ($t_1+t_2>0.05$). In the former case, the total time equals the nonlinear splitting time $t_1$ due to the absence of $U(t_2)$. While in the latter, positive gain appears only after $t_1+t_2>0.12$ when the fidelity between the instantaneous state and the initial state starts to increase. The largest gain appears at $t_1+t_2=0.18$, which is slightly after the instant of maximal fidelity occurring at $0.15$. Analogous delay is experimentally observed in the spin-1 BEC system~\cite{Liu2021}, which can be attributed to the imperfection of quasi-cyclic dynamics, with the output states remaining somewhat non-classical. 
We show in the Appendix~\ref{AppB} that the delay vanishes if the output state is fully disentangled, in which case the sensitivity will saturate QCRB.

\subsection{Scaling with particle number}\label{TNT5}
From the perspective of mean-field theory, the quasi-cyclic spin dynamics for an integrable system arises due to the closure of equal-energy contours, which remains in the limit of large particle numbers. Hence, we expect the cyclic nonlinear interferometers to take on qualitatively similar behavior regardless of system size. In Fig.~\ref{fig4}, we show the scaling of metrological gain as well as optimal $t_{1(2)}$ with particle number $N$. We focus on the largest metrological gain achievable in the non-Gaussian region by restricting $t_1$ and $t_1+t_2$ within $1.5T$, where the quasi-period $T$ denotes the instant when the fidelity with respect to the initial state returns to the maximum. In Fig.~\ref{fig4}(a), we find the cyclic nonlinear interferometer is capable of achieving measurement sensitivity following Heisenberg scaling of $\Delta\phi\propto N^{-1}$(blue circles). It is also worth noting that the deviation from the maximal QCRB over $t_1$ (black dotted line) remains $2.8~\rm{dB}$, making the protocol an efficient and close-to-optimal readout scheme for ENGS. In contrast, the linear readout protocol with $t_2=0$ follows a sub-Heisenberg scaling of $\Delta\phi\propto N^{-0.76}$ (white squares). 

Figure~\ref{fig4}(b) demonstrates the scaling of optimal spin dynamics durations $t_1$ and $t_1+t_2$. For all the explored atom numbers, these two values are slightly longer than, but follow the same scaling as the time for generating maximal QFI (blue dashed line) and the quasi-period (red dashed line). This result again reflects the important roles played by ENGS and cyclic spin dynamics in the nonlinear interferometers. 

\subsection{Robustness to detection noise}\label{TNT6}
\begin{figure}[t]
	\centering
	\includegraphics[width=1\linewidth]{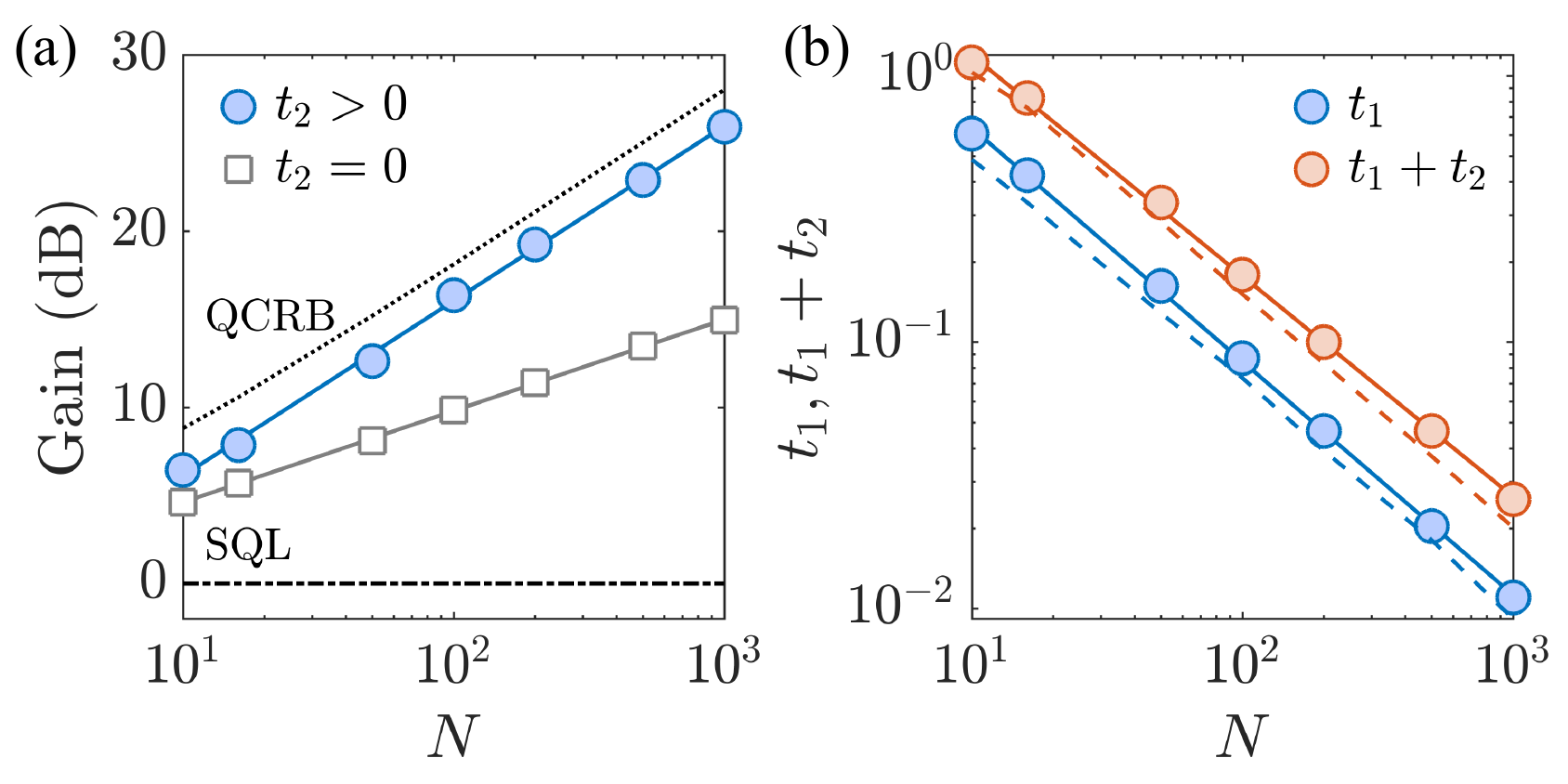}
	\caption{(a) Scaling of the metrological gains with atom number. Blue circles represent the gains optimized over $t_{1(2)}$. Gains achieved without nonlinear recombining ($t_2=0$) are shown by grey open squares for comparison. Black dotted and dash-dotted line denote the QCRB maximized over $t_1$ and SQL respectively. (b) Scaling of optimal $t_1$ and $t_1+t_2$. Blue (lower) and red (upper) dashed lines stand for the durations for local maxima of QCRB and the quasi-period $T$ respectively. We take $\Lambda=2$ and restrict $t_1$ (and $t_1+t_2$) to less than $1.5T$.}
 \label{fig4}
\end{figure}
Besides the effectiveness in readout of ENGS, the cyclic interferometry is also robust to detection noise. To show this, we model influence of detection noise with strength $\sigma$ as a Gaussian convolution of the initial positive operator valued measurement (POVM) $\{|m\rangle\langle m|\}$~\cite{IBR_spin}:
\begin{equation}
|\tilde{m}\rangle\langle\tilde{m}|=\sum_{m'}\Gamma_{m,m'}|m'\rangle\langle m'|,
\end{equation}
where 
\begin{equation}
\Gamma_{m,m'}=e^{-(m-m')^2/(2\sigma^2)}/\sum_me^{-(m-m')^2/(2\sigma^2)}.
\end{equation}
The projection of output state $\rho(\phi)$ on basis $|m\rangle$ then becomes $P_m(\phi)={\rm Tr}\left[\rho(\phi)|\tilde{m}\rangle\langle\tilde{m}|\right]$. To facilitate the calculation, we define the density matrix for the probe state as $\rho_p$, which is related to $\rho(\phi)$ through $$\rho(\phi)=R_mU(t_2)R_n(\phi)\rho_pR_n^{\dagger}(\phi)U^{\dagger}(t_2)R_m^{\dagger}.$$ Straight-forward derivations give
\begin{equation}
P_m(0)={\rm Tr}\left[|\tilde{m}\rangle\langle\tilde{m}|R_mU(t_2)\rho_pU^{\dagger}(t_2)R_m^{\dagger}\right],
\end{equation}
and
\begin{equation}
\begin{aligned}
&\left.\dfrac{\partial P_m(\phi)}{\partial\phi}\right\vert_{\phi=0}=\\
&-i{\rm Tr}\left[|\tilde{m}\rangle\langle\tilde{m}|R_mU(t_2)(\hat{S}_n\rho_p-\rho_p\hat{S}_n)U^{\dagger}(t_2)R_m^{\dagger}\right].
\end{aligned}
\end{equation}
We then find 
\begin{equation}
\langle\Delta\hat{S}_z\rangle^2=\sum_mm^2P_m(0)-\left(\sum_mmP_m(0)\right)^2,
\end{equation} and 
\begin{equation}
\left.\dfrac{\partial\langle\hat{S}_z(\phi)\rangle}{\partial\phi}\right|_{\phi=0}=\sum_mm\left.\dfrac{\partial P_m(\phi)}{\partial\phi}\right|_{\phi=0},
\end{equation} 
from which we obtain measurement precision in the presence of detection noise through error-propagation formula.

Figure~\ref{fig5} shows the numerical simulation results for $N=100$ particles. At each noise strength, we optimize $t_{1(2)}$ over the same range as in Fig.~\ref{fig4}. Compared to the linear readout scheme ($t_2=0$, squares), we find the cyclic nonlinear readout~($t_2>0$, circles) demonstrates an improved robustness by one order of magnitude. Metrological gain remains essentially unchanged until $\sigma=\sqrt{N}$, and sub-shot noise sensitivity can even be achieved when $\sigma$ is comparable to $N$. In contrast, linear readout scheme requires near single-particle resolution to maintain metrological gain beyond SQL.

We also compare the performance of the cyclic readout protocol with the conventional time-reversal one ($t_2<0$, triangles). Note that for a fair comparison between these two schemes, we have optimized both $t_1$ and $t_2$ in the time-reversal scheme (over the same range as in the cyclic scheme), instead of using the traditional time-reversal condition of $t_2=-t_1$~\cite{IBR_spin}. We find as shown in the figure, the optimal metrological gains as well as the robustness to detection noise in both protocols are quantitatively comparable.
\begin{figure}[t]
	\centering
	\includegraphics[width=0.95\linewidth]{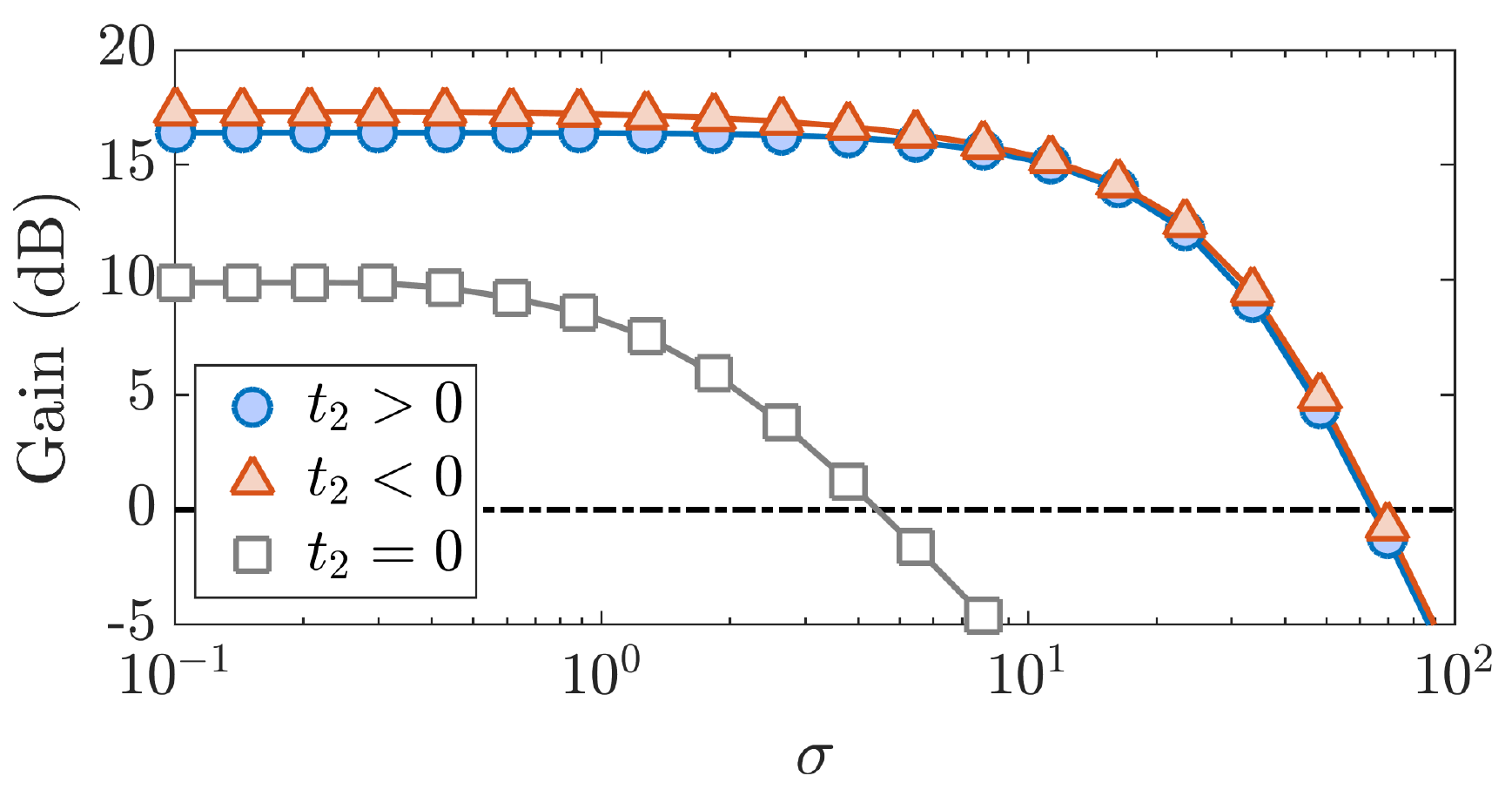}
	\caption{Robustness to detection noise for the TNT Hamiltonian. Nonlinear cyclic readout protocol (blue circles) demonstrates strong robustness compared to the linear readout scheme (grey squares), and comparable performance with time-reversed readout methods (red triangles). We take $N=100$ and optimize the gain over $t_1\in[0,T]$ and $|t_2|\in[0,1.5T]$, with $T$ being the quasi-period of the spin dynamics.}
 \label{fig5}
\end{figure}
\begin{figure*}[t]
	\centering
	\includegraphics[width=1\linewidth]{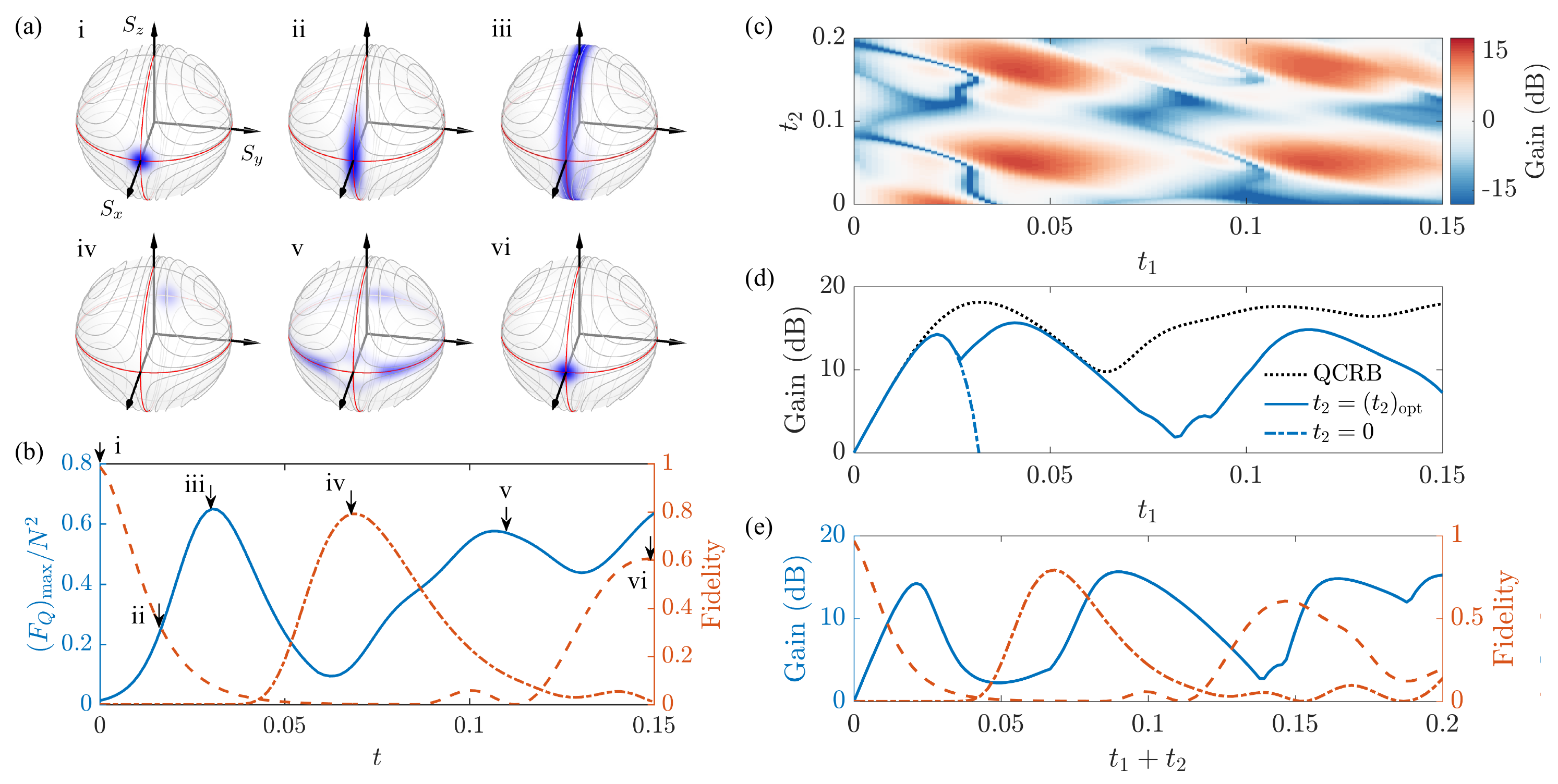}
	\caption{(a) Husimi representations of states generated by cyclic dynamics in the TACT model. Quantum noise of ENGS (ii, iii and iv) is refocused when the system evolves to either the initial state (vi) or its orthogonal state (iv), associated with the two unstable fixed points at $\langle\hat{\mathbf{S}}\rangle=(\pm N/2,0,0)$. Grey curves stand for mean field energy contours with separatrix highlighted by red lines. (b) Time-evolution of normalized QFI (blue solid line), fidelity with respect to the initial state $|s_x=N/2\rangle$ (red dashed line) or the opposite state $|s_x=-N/2\rangle$ (red dash-dotted line). (c) Dependence of metrological gain on $t_1$ and $t_2$. (d) Gains optimized over $t_2$ for given $t_1$ (blue solid line). Gains achieved without nonlinear readout (blue dash-dotted line) and QCRB (black dotted line) are also shown for comparison. (e) Optimal gain obtained at total time $t_1+t_2$ (blue line). Red dashed and dash-dotted lines show the same fidelity curves as in panel (b). $N=100$ is used in all panels.}
 \label{fig6}
\end{figure*}
\section{Cyclic nonlinear interferometry in the TACT model}\label{TACT}
High metrological gain achieved in the above cyclic nonlinear interferometer comes from quantum signal magnification, which arises as the system evolves back to the vicinity of the initial state. {Such a sequence of entangling-disentangling for quantum sensing is akin to the commonly investigated time-reversal-based quantum metrology}~\cite{Yurke:1986aa, Macri:2016aa, Davis, Frowis2016, Manceau:2017uq, Huang:2018ur, Huang:2018aa, TACT_magnification, Szigeti, Schulte2020ramsey, Hudelist2014, Linnemann:2016aa, spin_nematic, Burd, A.:2021aa, Hosten, Colombo:2022tz, lizeyang}. Although other studies~\cite{IBR_spin, Nolan, TNT_magnification} report that more flexible nonlinear readout beyond time-reversal can be designed to establish robust interferometry against detection noise, they are limited to consider phase estimation based on classical Fisher information, which could become experimentally cumbersome for many-body entangled states. Besides, it remains unclear if the explicit time-reversal (or tracing back to the initial state) is always required for quantum enhanced sensing based on low-order moment measurement of collective spin operators. 

In this section, we provide a negative answer to this question, and show the output state of nonlinear interferometry can be relaxed to a coherent spin state staying at the unstable fixed points in the absence of signal encoding. The familiar TACT Hamiltonian~\cite{squeezedmilestone, Kajtoch:2015aa, TACT_magnification} will be considered here, which is well-known for being capable of generating spin squeezing and ENGS, although it remains to be implemented experimentally. The original TACT Hamiltonian proposed by Kitagawa and Ueda~\cite{squeezedmilestone} takes the form of
\begin{equation}
\hat{\mathcal{H}}_{\rm{TACT}}=\dfrac{\chi}{2i}\left(\hat{S}_+^2-\hat{S}_-^2\right),
\end{equation}
where $\hat{S}_{\pm}=\hat{S}_x\pm i\hat{S}_y$. With a linear spin rotation $U=\exp[-i(\pi/2)\hat{S}_y]$, we obtain an equivalent Hamiltonian of (we take $\chi=1$ hereafter for brevity)
\begin{equation}
\begin{aligned}
\hat{H}_{\rm{TACT}}&=U\hat{\mathcal{H}}_{\rm{TACT}}U^{\dagger}\\
&=-(\hat{S}_y\hat{S}_z+\hat{S}_z\hat{S}_y).
\end{aligned}
\end{equation}
Different from the TNT model, the mean-field phase diagram of the TACT is characterized by two unstable fixed points located at $\langle\hat{\mathbf{S}}\rangle=(\pm N/2,0,0)$~\cite{Kajtoch:2015aa}. For dynamics starting from one of these points (for example,  $\langle\hat{S}_x\rangle=N/2$ as shown in Fig.~\ref{fig6}(a)i), the state initially flows along the separatrix in the $\hat{S}_x$-$\hat{S}_z$ plane, evolving from the Gaussian squeezed regime [Fig.~\ref{fig6}(a)ii] to the non-Gaussian regime [Fig.~\ref{fig6}(a)iii] and finally back to the vicinity of the initial state [Fig.~\ref{fig6}(a)vi] with a fidelity of $0.6$ at the end of a quasi-period [red dashed line in Fig.~\ref{fig6}(b)]. Specifically, we find at the instant of half quasi-period (marked by iv), the Husimi distribution converges to the opposite unstable fixed point at $\langle\hat{\mathbf{S}}\rangle=(-N/2,0,0)$, when the fidelity with respect to the corresponding coherent spin state reaches an even higher value of $0.8$ [red dash-dotted line in Fig.~\ref{fig6}(b)]. Since such a classical state is linked to the initial one simply through a spin rotation, the action of which is equivalent to the unitary transformation of the rotation $R_m$, one expects high metrological gain should also be achievable in this case.

We numerically verify this expectation in Fig.~\ref{fig6}(c), which shows a qualitatively different time dependence of the metrological gain from the TNT case. It is found that the non-Gaussian regime with positive gains is further divided into four small regions. This decomposition can be explained by the two local maxima of QFI and two unstable fixed points, the latter of which leads to two times of noise refocusing and signal magnification. More specifically, the nonlinear splitting time $t_1$ of the four peaks are located around the instants corresponding to the two local maxima of QCRB, as shown in Fig.~\ref{fig6}(d), while the optimal $t_2$ appears still right after the instants of noise refocusing where the fidelity with respect to either the initial state [red dashed line in Fig.~\ref{fig6}(e)] or its orthogonal classical state (red dash-dotted line) is maximized. Again, the largest obtainable gain of $15.6~\rm{dB}$ is close to the maximal QCRB of $18.1~\rm{dB}$, demonstrating the cyclic readout protocol is also nearly optimal in the TACT model.

Before conclusion, we want to stress that, the metrological performance of the TACT model is distinctively different from the previously reported nonlinear readout protocols, where the system is required to return to the initial state. 
Since the TACT model reaches the same level of measurement precision as the TNT but within half the total evolution time of the latter, it may suffer less from the decoherence and dissipation during the long-term nonlinear dynamics.

\section{Conclusion}\label{conclusion}
To summarize, we present a cyclic nonlinear interferometer in two-level spin systems that enables an efficient readout of ENGS. The general interferometry makes use of the intrinsic quasi-periodic spin dynamics to refocus quantum noise of probe state and magnify an encoded signal. As demonstrated in the TNT and TACT models, achievable sensitivities based on our protocols are found to nearly saturate the QCRB and follow Heisenberg-limited scaling with particle number. We also find in the TACT model, quantum magnification can be more generally realized when the state evolves to different classical state instead of the initial one. Compared to direct readout strategies based on measurements of the full probability distribution or high-order spin observables, our scheme is advantageous for parameter estimation since only measurements of mean value and fluctuation of population difference between the two levels are required. Its implementation avoids the challenging task of reversing many-body dynamics in nonlinear interferometry. Our work expands the repertoire of operational tools for ENGS, and can be readily applied in various physical systems such as two-component BEC~\cite{Strobel:2014ux}, cold atoms in cavity~\cite{Colombo:2022tz}, trapped ions~\cite{Bohnet:2016aa}, superconducting devices~\cite{Xu:2022aa}, dipolar gases with large spins~\cite{Chalopin:2018aa}, and Rydberg atom arrays~\cite{Omran:2019aa}, where ENGS has been  successfully generated by coherent spin dynamics.

\section*{Acknowledgements}
We thank F. Chen, X.-W. Li and Z.-Y. Hu for discussions.
This work is supported by the National Key R\&D
Program of China (Grant No. 2018YFA0306504),
the National Natural Science Foundation of China
(NSFC) (Grants Nos.  U1930201 and 92265205), and by the Innovation Program
for Quantum Science and Technology 
(2021ZD0302100).
M.X. is supported by the startup fund (Grant No. 1008-YAT21004) of NUAA.\par
~\\

\appendix
\section{Signal estimation based on higher-order spin operators}
Quasi-cyclic spin dynamics explored in this work does not perfectly disentangle the non-Gaussian probe state, which leads to the discrepancy of metrological gain from the QCRB. This deviation, however, can be mitigated through the measurement of higher-order collective spin operators. Here, we extend the pioneering work~\cite{Gessner:2019aa} by considering the nonlinear readout $U(t_2)$, and focus on the metrological gain at $\phi=0$.

For brevity, we define $\hat{\mathbf S}^{(k)}$ as the set containing all linear and also symmetric products of up to $k$ linear collective spin operators. Specifically, we have
\begin{widetext}
\begin{equation}
\hat{\mathbf S}^{(1)}=\left(\hat{S}_x, \hat{S}_y, \hat{S}_z\right),
\end{equation}
\begin{equation}
\hat{\mathbf S}^{(2)}=\left(\hat{S}_x, \hat{S}_y, \hat{S}_z, \hat{S}_x^2, \hat{S}_y^2, \hat{S}_z^2,\{\hat{S}_x,\hat{S}_y\}/2,\{\hat{S}_x,\hat{S}_z\}/2,\{\hat{S}_y,\hat{S}_z\}/2\right),
\end{equation}  
\begin{equation}
\begin{aligned}
\hat{\mathbf S}^{(3)}=&\bigg(\hat{S}_x, \hat{S}_y, \hat{S}_z, \hat{S}_x^2, \hat{S}_y^2, \hat{S}_z^2, \hat{S}_x^3, \hat{S}_y^3, \hat{S}_z^3,\{\hat{S}_x,\hat{S}_y\}/2, \{\hat{S}_y,\hat{S}_z\}/2, \{\hat{S}_x,\hat{S}_z\}/2,\\
&(\hat{S}_x^2\hat{S}_y+\hat{S}_x\hat{S}_y\hat{S}_x+\hat{S}_y\hat{S}_x^2)/3, (\hat{S}_x^2\hat{S}_z+\hat{S}_x\hat{S}_z\hat{S}_x+\hat{S}_z\hat{S}_x^2)/3,(\hat{S}_y^2\hat{S}_x+\hat{S}_y\hat{S}_x\hat{S}_y+\hat{S}_x\hat{S}_y^2)/3, \\
&(\hat{S}_y^2\hat{S}_z+\hat{S}_y\hat{S}_z\hat{S}_y+\hat{S}_z\hat{S}_y^2)/3,(\hat{S}_z^2\hat{S}_x+\hat{S}_z\hat{S}_x\hat{S}_z+\hat{S}_x\hat{S}_z^2)/3, (\hat{S}_z^2\hat{S}_y+\hat{S}_z\hat{S}_y\hat{S}_z+\hat{S}_y\hat{S}_z^2)/3,\\
&(\hat{S}_x\hat{S}_y\hat{S}_z+\hat{S}_x\hat{S}_z\hat{S}_y+\hat{S}_y\hat{S}_z\hat{S}_x+\hat{S}_y\hat{S}_x\hat{S}_z+\hat{S}_z\hat{S}_x\hat{S}_y+\hat{S}_z\hat{S}_y\hat{S}_x)/6\bigg).
\end{aligned}
\end{equation}  
\end{widetext}
To determine the optimal measurement precision, we follow the same spirit as in the main text to define matrices $M$ and $Q$, with the dimension decided by the number of elements in the $\hat{\mathbf{S}}^{(k)}$ set. The sensitivity $(\Delta\phi)^{-2}$ then remains upper bounded by Eq.~\eqref{optimal_sensitivity}. However, since the dimension of $\mathbf{n}$ remains to be 3, the maximal value of the right-hand-side is given by the maximal eigenvalue of the $3\times3$ submatrix which has the same first three columns and rows as the momentum matrix $MQ^{-1}M^T$~\cite{Gessner:2019aa}.

Figure~\ref{fig_appA} shows the calculation results with TNT model. We see with the increment of order $k$, the metrological gain at $t_1\approx0.05$ improves, and QCRB can be saturated within a wider range of $t_1$. It is also noted that the optimal gain of 16.4 dB achieved by measuring $\mathbf{\hat{S}}^{(1)}$ after quasi-cyclic dynamics (blue solid line) is comparable to the optimal value of 16.6 dB obtained through direct measurements of $\mathbf{\hat{S}}^{(3)}$ (red dashed line), which thus highlights the efficiency of our protocol in reading out ENGS.
\begin{figure}[t]
	\centering
	\includegraphics[width=1\linewidth]{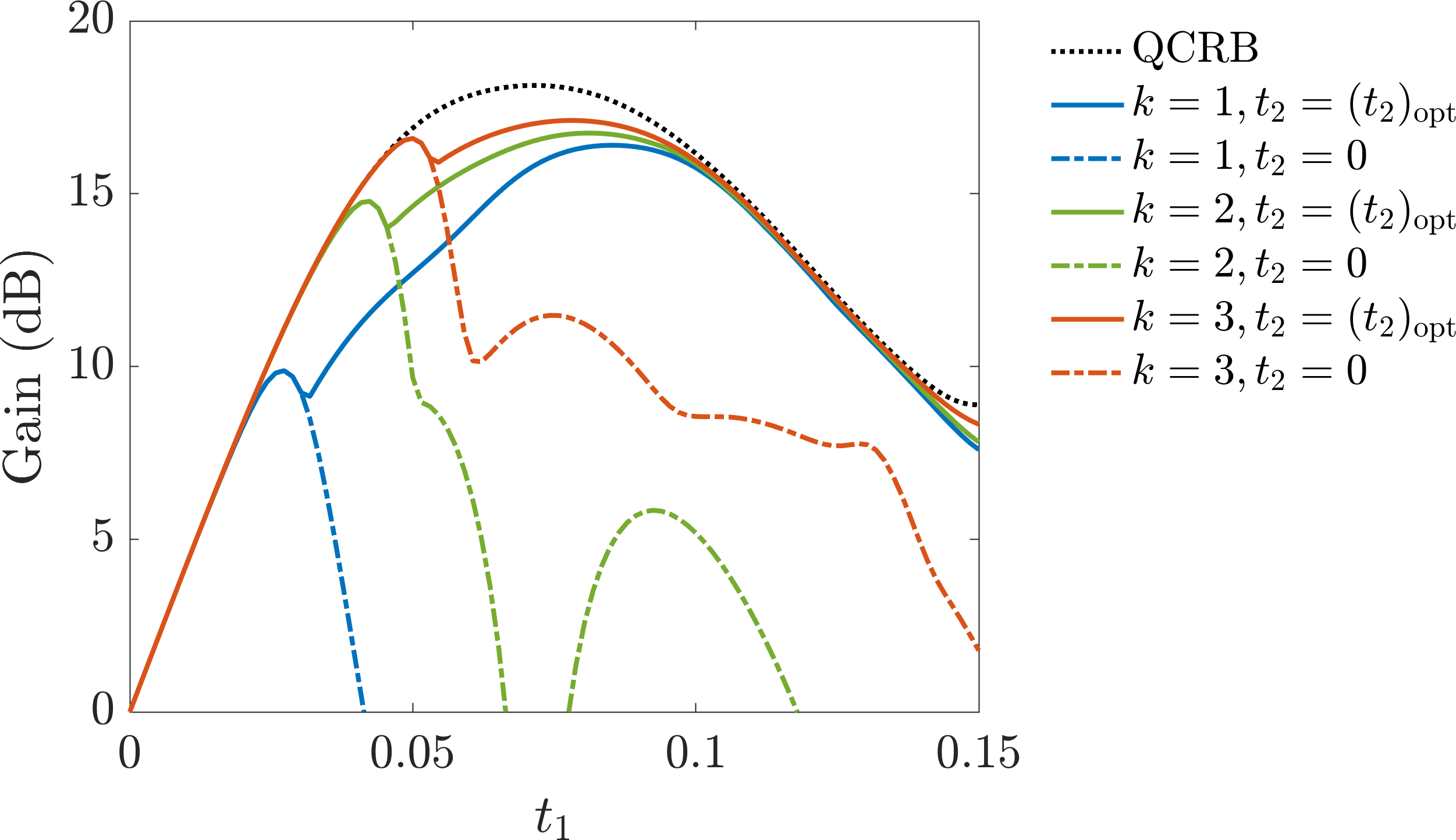}
	\caption{Metrological gain based on measuring higher-order collective spin operators. Blue, green and red solid (dashed) lines denote the gain obtained by measuring $\mathbf{\hat{S}}^{(1)}$, $\mathbf{\hat{S}}^{(2)}$ and $\mathbf{\hat{S}}^{(3)}$ respectively with $t_2$ optimized over $[0,0.2]$ (fixed to be 0). The simulations are carried out with $N=100$, $\chi=1$ and $\Lambda=2$ in the TNT model.}
 \label{fig_appA}
\end{figure}

\label{AppA}

\section{Quantum magnification}
It is not generally guaranteed QCRB can be saturated with measurements of mean values and standard deviations of linear spin operators. However, for the special case where the output states of nonlinear interferometry are classical states, we show here that the above lower-moments measurement is sufficient to saturate QCRB. In this case, the instant of maximal metrological gain is the same as the one of optimal disentanglement (or the fidelity with respect to a coherent spin state reaches $1$), and the time delay in Fig.~\ref{fig3}(c) disappears. 

We denote the probe state as $|\psi_p\rangle$ and signal encoding operator as $\hat{S}_p$. The state after encoding a signal of $\phi$ then becomes $|\psi_p'\rangle=e^{i\phi\hat{S}_p}|\psi_p\rangle$. In the limit of $\phi\rightarrow0$, we can expand the expression for fidelity between these two states in Taylor series:
\begin{equation}
\begin{aligned}
F(\phi)&=|\langle\psi_p|e^{i\phi\hat{S}_p}|\psi_p\rangle|^2\\
&=1-\phi^2(\Delta\hat{S}_p)^2+O(\phi^4),
\end{aligned}
\end{equation}
where $(\Delta\hat{S}_p)^2=\langle\psi_p|\hat{S}_p^2|\psi_p\rangle-(\langle\psi_p|\hat{S}_p|\psi_p\rangle)^2$ gives the QFI of $F_Q=4(\Delta\hat{S}_p)^2$. Since the fidelity is a constant during unitary evolution, we have the relationship
\begin{equation}
|\langle\psi_o'|\psi_o\rangle|^2=1-{\phi^2}F_Q/4+O(\phi^4),
\end{equation}
with $|\psi_o\rangle$ and $|\psi_o'\rangle$ the corresponding output states for without and with signal encoding respectively. For coherent output states, $|\psi_o\rangle$ and $|\psi_o'\rangle$ are related to each other through a linear spin rotation, namely $
|\psi_o'\rangle=e^{i\theta\hat{S}_o}|\psi_o\rangle$,
which gives 
\begin{equation}
\begin{aligned}
|\langle\psi_o'|\psi_o\rangle|^2&=1-\theta^2(\Delta\hat{S}_o)^2+O(\theta^4)\\
&=1-{\theta^2}N/4+O(\theta^4),
\end{aligned}
\end{equation}
where $(\Delta\hat{S}_o)^2=N/4$ comes from the projection noise of a coherent state.
Comparing the above two equations, we then have $\theta=\phi\sqrt{F_Q/N}$ with $\sqrt{F_Q/N}$ being the magnification factor of the spin rotation. Since the measurement precision of $\theta$ with respect to a coherent state is lower bounded by the SQL $\Delta\theta=1/\sqrt{N}$, which can be accessed with linear measurement, we finally have $\Delta\phi=1/\sqrt{F_Q}$, or the QCRB, for the sensitivity of nonlinear interferometry.
\label{AppB}

\bibliography{quantum_metrology}


\clearpage

\end{document}